\documentclass[useAMS,usenatbib]{mn2e}  
\usepackage{graphicx}
\usepackage{txfonts}
\newcommand \avbparr{$\langle B_{\|}\rangle$}
\newcommand \sigmaRM{$\sigma_{\mathrm{RM}}$}
\newcommand \sigmaAvbparr{$\sigma_{\langle B_{\|}\rangle}$}
\newcommand \sigmaDMinf{$\sigma_{\mathrm{DM}_{\infty}}$}

\newcommand \DMinf{DM$_{\mathrm{\infty}}$}
\newcommand \DMhalf{DM$_{1/2}$}
\newcommand \avdist{$\langle \mathrm{dist}\rangle_{n_{\mathrm{e}}}$}
\newcommand \chitwored{$\chi^2_{\mathrm{red}}$}

\begin{document}
   \title[Latitude dependence of RMs of NVSS sources]{The latitude dependence of the rotation measures of NVSS sources}

   \author[D.H.F.M. Schnitzeler]{D.H.F.M. Schnitzeler$^1$\thanks{E-mail: Dominic.Schnitzeler@csiro.au}\\
           $^1$ Australia Telescope National Facility, CSIRO Astronomy and Space Science, Marsfield, NSW 2122, Australia
          }

   \date{Accepted 2010 September 16. Received 2010 August 23; in original form 2010 April 22}

   \pagerange{}\pubyear{2010}

   \maketitle

   \label{firstpage}

   \begin{abstract}
     In this Letter I use the variation of the spread in rotation measure (RM) with Galactic latitude to separate the Galactic from the extragalactic contributions to RM. This is possible since the latter does not depend on Galactic latitude. As input data I use RMs from the catalogue by Taylor, Stil, and Sunstrum, supplemented with published values for the spread in RM (`\sigmaRM') in specific regions on the sky. I test 4 models of the free electron column density (which I will abbreviate to `\DMinf') of the Milky Way, and the best model builds up \DMinf\ on a characteristic scale of a few kpc from the Sun. \sigmaRM\ correlates well with \DMinf. The measured \sigmaRM\ can be modelled as a Galactic contribution, consisting of a term $\sigma_{\mathrm{RM,MW}}$ that is amplified at smaller Galactic latitudes as 1/sin$|b|$, in a similar way to \DMinf, and an extragalactic contribution, $\sigma_{\mathrm{RM,EG}}$, that is independent of latitude. This model is sensitive to the relative magnitudes of $\sigma_{\mathrm{RM,MW}}$ and $\sigma_{\mathrm{RM,EG}}$, and the best fit is produced by $\sigma_{\mathrm{RM,MW}}$ $\approx$ 8 rad/m$^2$ and $\sigma_{\mathrm{RM,EG}}$ $\approx$ 6 rad/m$^2$. The 4 published values for \sigmaRM\ as a function of latitude suggest an even larger $\sigma_{\mathrm{RM,MW}}$ contribution and a smaller $\sigma_{\mathrm{RM,EG}}$. 
This result from the NVSS RMs and published \sigmaRM\ shows that the Galactic contribution dominates structure in RM on scales between about 1\degr\ -- 10\degr\ on the sky. I work out which factors contribute to the variation of \sigmaRM\ with Galactic latitude, and show that the $\sigma_{\mathrm{RM,EG}}$ I derived is an upper limit. Furthermore, to explain the modelled $\sigma_{\mathrm{RM,MW}}$ requires that structure in \avbparr\ has a 1-$\sigma$ spread $\lesssim$ 0.4 $\mu$G.
   \end{abstract}


   \begin{keywords}
   polarization -- Galaxy: general -- ISM: magnetic fields -- galaxies: magnetic fields
   \end{keywords}

%

\section{Introduction}
In recent years, models of the large-scale Galactic magnetic field
have increased greatly in both complexity and breadth of input data
and output variables (see e.g. \citealt{fauvet10}, \citealt{jaffe10},
\citealt{nota10}, \citealt{jansson09}, \citealt{waelkens09},
\citealt{sun09}, \citealt{men08}, \citealt{sun08},
\citealt{brown07}). One popular approach to determine magnetic field
strengths is to look for Faraday rotation of the polarized emission
coming from extragalactic sources (see e.g. \citealt{gaensler05}).
The rotation measure (RM) quantifies the amount of Faraday rotation
between the source of the emission and us, the observers, and it
depends on the free electron density, $n_{\mathrm{e}}$, and the
magnetic field component along the line of sight, $B_\|$, that the
emission encounters along its path:

$$ \mathrm{RM}\ [\mathrm{rad/m}^2]\ = 0.81\int_{\mathrm{source}}^{\mathrm{observer}} n_{\mathrm{e}}\ [\mathrm{cm}^{-3}]\ 
B_\|\ [\mu\mathrm{G}]\ \mbox{d}l\ [\mathrm{pc}] $$

\noindent 
where d$l$ an infinitesimal part of the line of sight towards the observer. 
The RM can tell us about the magnetic field strength along the line of
sight, when the electron density contribution to RM is accounted for,
for example by determining the free electron column density towards
the source of the emission, which is known as the dispersion measure
(DM) in the pulsar community.

In this Letter I will focus on one particular observational aspect,
which is that the width of the RM distribution (\sigmaRM) of
extragalactic sources increases closer to the Galactic plane. This
increase furthermore closely matches the increase in DM closer to the
plane. I will use this, in combination with the fact that
the extragalactic RM contribution is independent of Galactic latitude,
to separate this component from the Galactic RM contribution. This
Letter is organised as follows. First I will describe the NVSS RM
catalogue by \citet{taylor09} that I use for my analysis in
Sect. \ref{Sect.: the data}, and in Sect. \ref{Sect.: RM_lat} I
describe the analysis that led to the observation that \sigmaRM\
depends on Galactic latitude.
In Sect. \ref{Sect.: ne_model} I determine how well 4 models for the
free electron density in the Galaxy predict the observed DM of pulsars
at known distances.
Finally, in Sect. \ref{Sect.: ratio} I will model the Galactic and
extragalactic contributions to \sigmaRM, and separate the two.
Throughout this Letter I have calculated statistics in a way that is
robust against outliers.



\section{The data}\label{Sect.: the data}
%
%
\citet{taylor09} determined RMs
for 37.543 sources from the NVSS catalogue, which covers declinations
$>$ -40\degr. 
This means that the NVSS sources are separated on average by
about 1\degr\ on the sky. Even though the accuracy of the individual
RMs is limited (the median error in RM is about 11 rad/m$^2$), the
sheer size of the NVSS RM catalogue makes it very useful for studying
the properties of the Galactic magnetized interstellar medium on large
scales.  In Fig. \ref{b_rm.fig} I show the distribution of NVSS RMs as
a function of Galactic latitude. 

Close to the Galactic plane, the RMs are much higher than further
away, and this impacts the reliability of NVSS RMs at small Galactic
latitudes. First, a large RM will induce a large amount of bandwidth
depolarization, which reduces the polarized signal/noise ratio. Also,
since the NVSS uses only 2 frequency bands, large RMs will suffer from
n$\pi$ ambiguities due to the periodicity of the polarization angles,
and they will show up in the NVSS RMs as small RMs. Taylor et al. have
suppressed the latter effect when they fitted RMs by also estimating
the amount of bandwidth depolarization that would be expected for
large RMs. Because of these reasons, I only use NVSS sources with
$|$RM$|$ $<$ 300 rad/m$^2$ and Galactic latitudes $|b|$ $\ge$ 20\degr\
in my analysis. The NVSS catalogue is not very sensitive to diffuse
Galactic structure, which would have introduced further depolarization
effects.



\section{The spread in RM as a function of latitude}\label{Sect.: RM_lat}
\begin{figure}
\resizebox{\hsize}{!}{\includegraphics{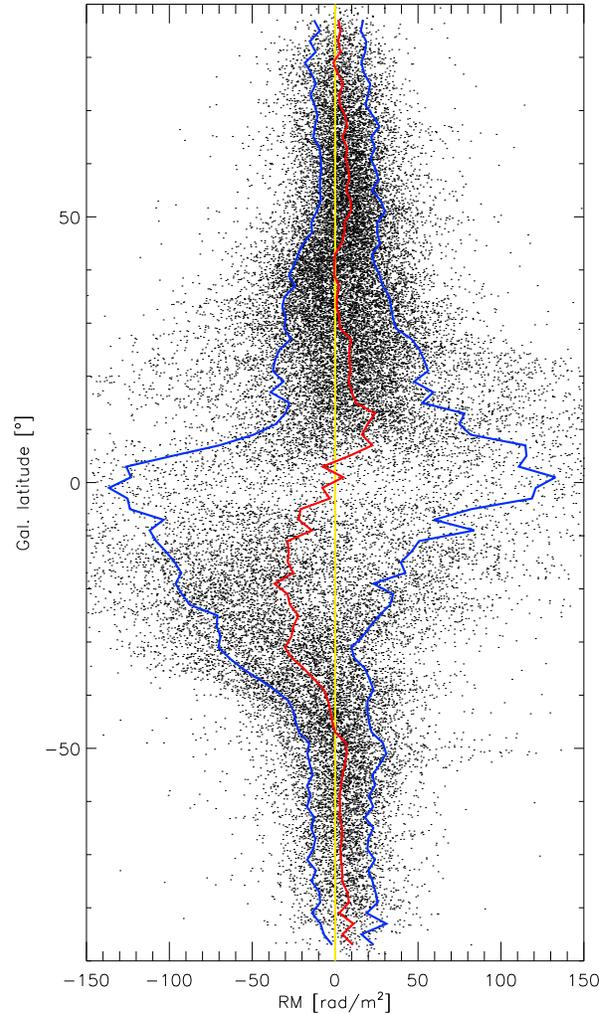}}

\caption{Distribution of NVSS RMs as a function of Galactic latitude, which clearly shows the broadening of the RM distribution closer to the Galactic plane. The average RM and 1-$\sigma$ spread around the average RM are calculated for 2\degr\ bins in Galactic latitude, and are shown as the red line and the blue lines on either side of the red line. Only lines of sight with $|$RM$|$ $<$ 300 rad/m$^2$ are included in this and the following figures; 3\% of the lines of sight have $|$RM$|$ $>$ 150 rad/m$^2$ and are not shown in this figure.
}
\label{b_rm.fig}
\end{figure}

Fig. \ref{b_rm.fig} shows that the spread in the NVSS RM increases for
lines of sight closer to the Galactic plane, and in this Letter I will
show that this increase very closely follows the increase in the
Galactic \DMinf. Fig. 5 from \citet{taylor09} also shows this increase
in \sigmaRM. The NVSS RMs should however be corrected for 2 effects
before this conclusion can be drawn.

First, the average RM will vary in a strip along Galactic longitude,
an effect which becomes more pronounced close to the Galactic plane,
and this increases \sigmaRM\ when left uncorrected. I divided each
strip along Galactic latitude into bins, and determined the average RM
for each bin. For latitudes within 77\degr\ of the plane bins are
5\degr/cos($b$) $\times$ 4\degr\ in size ($\Delta l \times \Delta
b$). The 1/cos($b$) dependence of the bin width produces bins with a
constant surface area on the sky, which guarantees that each bin will
contain about 20 NVSS RMs, and that the maximum separation between two
points in a bin is constant. Doubling $\Delta l$ does not
significantly influence the results from Sect. \ref{Sect.: ratio}.
I divided the cap regions above $|b|$ = 77\degr\ into intervals with
$\Delta l$ = 20\degr\ that converge at $b$=$\pm$ 90\degr. I then
calculated a cubic spline through the bin-averaged RM at the
longitudes of the individual NVSS RMs, and subtracted this spline fit
from the RMs. This removes structure in RM on scales larger than
10\degr\ (Nyquist sampling; for bins with $|b|$ $<$ 77\degr).  The
resulting RM distribution for each strip along Galactic latitude is
much better centred on 0 rad/m$^2$ than the uncorrected measurements,
and variations in $\langle$RM$\rangle$ that would increase \sigmaRM\
are strongly reduced.

Second, the \sigmaRM\ that were determined from the corrected NVSS RMs
should be corrected in a statistical sense for the measurement errors
of the individual NVSS RMs. I estimate the \sigmaRM\ that is produced
by just the variation in uncertainty in RM of the NVSS sources
(`errRM') by using a Monte Carlo simulation, which I set up as
follows.  First, I randomly draw errRM for 1000 lines of sight in the
NVSS catalogue, and I then use a random number generator\footnote{The
  $\mathtt{randomn}$ generator in \small{IDL}, which is similar to the
  $\mathtt{ran1}$ generator from Sect. 7.1 in \citet{numrecipes} } to
find a distribution of RM based on these 1000 errRM. I then calculate
the \sigmaRM\ of this distribution, and repeat this process 10$^5$
times. The total number of independently drawn random variables for
these 1000 lines of sight therefore does not exceed 10$^8$, the
maximum number of independent draws for the random number generator
that I used. I repeated this process for 35 sets of 1000 lines of
sight, to cover most of the errRM with $|b|$ $>$ 5\degr. An analysis
of the \sigmaRM\ found after each run shows that the spread due to the
errors in NVSS RMs gives a $\sigma_{\mathrm{errRM}}$ = 10.4 $\pm$ 0.4
rad/m$^2$.
%

%
%

\begin{figure}
\resizebox{\hsize}{!}{\includegraphics{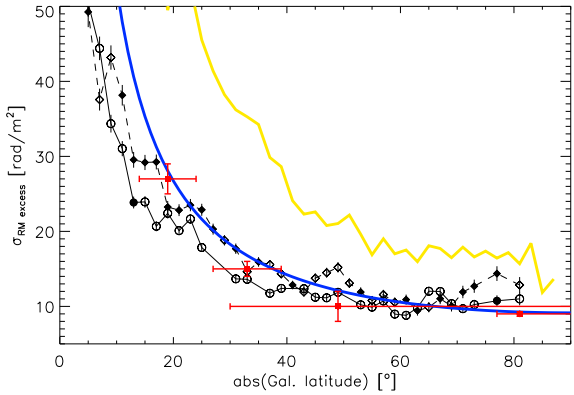}}

\caption{Distribution of the excess \sigmaRM\ as a function of Galactic latitude; circles (connected by a solid line) and squares (connected by a dashed line) indicate positive and negative Galactic latitudes respectively. The data points at Galactic latitudes of $\pm$1\degr\ and $\pm$ 3\degr\ are missing, their \sigmaRM\ values are 117 rad/m$^2$ and 85 rad/m$^2$ resp. The plotted data points contain at least 15 RMs, and their RM distribution can be fitted by a Gaussian with a reduced $\chi^2$ $<$ 2 (filled symbols) or between 2 $<$ \chitwored\ $<$ 4 (open symbols). The error bars (which are often smaller than the plot symbols) are calculated as the error in the mean RM of each latitude bin. Also shown are the uncorrected \sigmaRM\ calculated from Fig. \ref{b_rm.fig} (yellow line; sampled in 2\degr\ bins, and averaged over positive and negative latitudes) and the \sigmaRM\ (not the excess \sigmaRM!) from Table \ref{sigmaRM.table} (red symbols). The red horizontal error bars indicate the range in Galactic latitude that is covered by these data sets.
To illustrate how well the \DMinf\ distribution traces the excess \sigmaRM\ at $|b|$ $>$ 20\degr, I plot this distribution after scaling it by a factor of 0.37 (blue line; see Sect. \ref{Sect.: ne_model}). 
}
\label{sigma_rm.fig}
\end{figure}

For 2 Gaussians, the widths ($\sigma$) add quadratically, and
$\sigma^2_{\mathrm{errRM\ +\ nature}}$ =
$\sigma^2_{\mathrm{errRM}}+\sigma^2_{\mathrm{nature}}$.
The `excess \sigmaRM', $\sigma_{\mathrm{nature}}$, can then be
recovered from the $\sigma_{\mathrm{errRM\ +\ nature}}$ measured with
NVSS, and the $\sigma_{\mathrm{errRM}}$ that I simulated with a Monte
Carlo process. A second Monte Carlo simulation showed that this
relation is not exactly correct due to a slight non-Gaussianity in the
distribution of the RM errors in NVSS.
This shows up as a 6\% difference between the expected \sigmaRM\
excess and the measured \sigmaRM\ excess, if the latter is 10
rad/m$^2$ (this difference decreases with increasing \sigmaRM\
excess). I corrected the \sigmaRM\ excess for this effect, and I will
use these corrected \sigmaRM\ in the remainder of the Letter.



In Fig. \ref{sigma_rm.fig} I plot the distribution of the excess
\sigmaRM\ as a function of Galactic latitude for both positive (solid
line) and negative (dashed line) latitudes; in
Fig. \ref{sigma_avbparr.fig} I will distinguish between positive and
negative latitudes in the same way.  I also show the \sigmaRM\
distribution of the uncorrected NVSS RM data (yellow line). It is
striking that there is a gap in \sigmaRM\ between positive and
negative latitudes of about 2 rad/m$^2$, but its origin is unknown to
me. \citet{taylor09} and \citet{mao10} noticed a similar
  RM difference between the north and south Galactic caps.

I compiled a list of published values of \sigmaRM\ in different
regions on the sky, which I show in Table \ref{sigmaRM.table}, and
which I also include in Figs. \ref{sigma_rm.fig} --
\ref{sigma_avbparr.fig}.  These data sets are not limited by having
only 2 frequency bands like the NVSS RMs.  \citet{feain09}
  removed a large-scale RM gradient from their data before calculating
  \sigmaRM, and \citet{mao10} found an (almost) flat power spectrum
  for their Galactic cap regions, so their \sigmaRM\ is not
  appreciably increased by a variation in RM. \citet{melanie04} argue
  that Galactic RM contributions are unimportant at latitudes $|b|$
  $>$ 30\degr, and they would therefore not increase \sigmaRM. It
  turns out that there is some variation in Galactic RM even at these
  latitudes, which means that their \sigmaRM\ is an upper limit to the
  excess \sigmaRM. \citet{gaensler05} did not correct for a smooth
  variation in RM, and also their \sigmaRM\ is therefore an upper
  limit. A variation in RM only increases \sigmaRM, but since the
  \sigmaRM\ that \citet{melanie04} and \citet{gaensler05} calculate
  are very similar to the excess \sigmaRM\ that I calculate from the
  NVSS RMs at these latitudes, their \sigmaRM\ can be considered as
  tight upper limits.  Since the data sets from Table
\ref{sigmaRM.table} have (very) small measurement errors in RM, I will
use the \sigmaRM, not the excess \sigmaRM, of these data sets.
%
%
Since the data from \citet{melanie04} and the polar cap regions from
\citet{mao10} span a wide range in Galactic latitudes, I plotted these
symbols at their surface-area-weighted average Galactic
latitudes\footnote{I define the surface-area averaged Galactic
  latitude as\newline $\langle b\rangle = \frac{ \int\limits_{b\ >\
      b_{\mathrm{min}}}^{90^{\circ}}\mathrm{area\ in\ annulus}\
    \times\ b\ \mathrm{d}b} {\int\limits_{b\ >\
      b_{\mathrm{min}}}^{90^{\circ}}\mathrm{area\ in\ annulus}\
    \mathrm{d}b} $ for data with $b$ $>$ $b_{\mathrm{min}}$.}, which
are $\langle b\rangle$ = 49\degr\ and $\langle b\rangle$ = 81\degr\
respectively. A uniform source density will result in more sources at
smaller Galactic latitudes, so using the surface area to weigh the
Galactic latitudes reflects a weighting with the number of sources in
each annulus. 

The correction for the variation in $\langle$RM$\rangle$ of the NVSS
RMs is not perfect, and the RM distribution in the latitude bins often
cannot be properly fitted by a Gaussian distribution, which is
reflected in the high values for the reduced $\chi^2$ (\chitwored) of
the Gaussian fits. 
However, Figs. \ref{sigma_rm.fig} and \ref{sigma_avbparr.fig} show
that there is a reasonable agreement between the \sigmaRM\ from Table
\ref{sigmaRM.table}, and the excess \sigmaRM\ that I calculated, an
indication that the corrections that I applied to the \sigmaRM\ of the
NVSS sources work well. Furthermore, latitude bins with \chitwored\
between 2 -- 4 show the same global behaviour with latitude as the
bins with \chitwored\ $<$ 2. For my analysis in the remainder of this
Letter I will therefore use the excess \sigmaRM\ of the NVSS sources
with \chitwored\ $<$ 4, and the \sigmaRM\ from the literature.  I will
distinguish excess \sigmaRM\ from NVSS with \chitwored\ $<$ 2 from
those with 2 $<$ \chitwored\ $<$ 4 by using filled symbols for the
former, and open symbols for the latter.


\begin{table}
\centering
\begin{tabular}{cr@{\degr}lr@{$\pm$}lc}
\hline
region & \multicolumn{2}{c}{$\langle b\rangle$} & \multicolumn{2}{c}{\sigmaRM\ [rad/m$^2$]} & source\\ \hline
Centaurus A       & 19 & & \hspace*{4mm}27 & 2$^{\mathrm{a}}$ & \citet{feain09} \\
LMC               & -33 & & 15 & 1$^{\mathrm{a}}$ & \citet{gaensler05}$^{\mathrm{b}}$ \\
$|b|$ $>$ 30\degr & 49 &$^{\mathrm{c}}$  & 10 & 2 & \citet{melanie04}$^{\mathrm{d}}$ \\
$|b|$ $>$ 77\degr & 81 &$^{\mathrm{c}}$  &  9 & 0.3$^{\mathrm{a}}$ & \citet{mao10} \\
\hline
\end{tabular}
\caption{Values of \sigmaRM\ compiled from the literature.}
\begin{list}{}{}
\item[$^{\mathrm{a}}$]: calculated as the standard error in the mean, using the number of sources that the authors listed to lie outside the target object.
\item[$^{\mathrm{b}}$]: and Ann Mao, private communication
\item[$^{\mathrm{c}}$]: $\langle b\rangle$ calculated as the surface-area-weighted average Galactic latitude of that data set
\item[$^{\mathrm{d}}$]: and Melanie Johnston-Hollitt, private communication
\end{list}
\label{sigmaRM.table}
\end{table}

\section{A model for the free electron density}\label{Sect.: ne_model}
Several models have been proposed in the literature to describe the
free electron density in the Milky Way. Here I test the accuracy of
the NE2001 model by \citet{cordeslazio02}\footnote{and their website,
  http://rsd-www.nrl.navy.mil/7213/lazio/ne\_model/}, the model by
\citet{gaensler08}, and the model by \citet{bm08}, by comparing the
predicted DM to the measured DM for 65 pulsars of which the distance
is known to within 33\%. I compiled this sample from the ATNF pulsar
catalogue (\citealt{manchester05}), downloaded from this
website\footnote{http://www.atnf.csiro.au/research/pulsar/psrcat/} on
25/05/2009, and the papers by \citet{chatterjee09} and
\citet{deller09}.  I also fitted a plane-parallel exponential model of
the free electron density to this data set, taking into account the
uncertainty in the pulsar distances. This model uses a more recent
version of the catalogues that Gaensler et al. and Berkhuijsen et
al. compiled, it includes also pulsars at latitudes within 40\degr\
from the Galactic plane (the model by Gaensler et al. is constrained
by pulsars at latitudes $|b|$ $>$ 40\degr), and it puts a stronger
constraint on which pulsars to use (distance errors $<$ 33\%, where
Berkhuijsen et al. required $<$ 50\%). The best fitting model has a
mid-plane electron density $n_{\mathrm{e,0}}$ = 0.02 $\pm$ 0.0001
cm$^{-3}$ and scale height $h$ = 1.225 $\pm$ 0.007 kpc. (see also
Schnitzeler \& Katgert, in preparation)  The ratios of the predicted
DM/observed DM for these four models are 1.22 $\pm$ 0.53, 0.8 $\pm$
0.31, 1.05 $\pm$ 0.41 and 0.99 $\pm$ 0.37 resp. (mean $\pm$ 1-$\sigma$
spread; the error in the mean is smaller by a factor of $\sqrt{65}
\approx$ 8), so I selected the final model to work with.
The 65 pulsars that I selected cover a wide range of heights above the
Galactic midplane, out to about 10 kpc, so the fitted scale height is
well sampled by this pulsar distribution.

The 3 models that use a plane-parallel exponential profile for the
free electron density in the Milky Way (Gaensler et al., Berkhuijsen
et al., and the model I fitted) differ in mid-plane density and scale
height. However, when integrating these models out to infinity (a good
approximation for the NVSS sources), all 3 models should show the same
asymptotic behaviour for a given Galactic latitude $b$: \DMinf\ =
24.4/sin$|b|$. DM($b$=$\pm$90\degr) = 24.4 cm$^{-3}$pc is the median
DM that I calculated for the 8 pulsars from the sample that lie
further than 4 kpc from the Galactic plane. (\citealt{bm08} find
\DMinf($b$=$\pm$90\degr) = 21.7cm$^{-3}$pc, and \citealt{gaensler08}
find \DMinf($b$=$\pm$90\degr) = 25.6 cm$^{-3}$pc) In
Fig. \ref{sigma_rm.fig} I overplot the \DMinf\ predicted by this
model. To emphasize how well \DMinf\ follows the trend in \sigmaRM\
excess, I scaled the \DMinf\ by a factor of 0.37, which is the
  average value of \sigmaRM\ $\times$ sin$|b|$/24.4 for NVSS RMs at
  $|b|$ $\ge$ 20\degr.  The resulting match in Fig. \ref{sigma_rm.fig}
  is not perfect, since the values of \sigmaRM\ $\times$ sin$|b|$ from
  Fig. \ref{sigma_avbparr.fig} depend on latitude. In
  Sect. \ref{Sect.: ratio} I will show that the match can be improved
  by taking the $b$-independent contribution from
  $\sigma_{\mathrm{RM,EG}}$ to \sigmaRM\ into account.



To determine the characteristic distance over which most of the \DMinf\
is built up, I introduce the electron-density weighted average
distance along the line of sight, \avdist\footnote{I define the
  electron-density weighted average distance, \avdist, as \newline
  \avdist=$\frac{\int\limits_{0}^{\infty} l n_{\mathrm{e}}\
    \mbox{d}l}{\int\limits_{0}^{\infty} n_{\mathrm{e}}\ \mbox{d}l}$,
  where both integrals are along the line of sight. For a free
  electron density that decreases exponentially away from the Galactic
  plane, \avdist\ = $h$/sin$|b|$} = $h$/sin$|b|$. Note that \avdist\
is somewhat larger than the distance over which half the \DMinf\ is
built up, \DMhalf=ln(2)$\times\ h$/sin$|b|$ $\approx$ 0.7
\avdist. When going from $|b|$=90\degr\ to $|b|$=20\degr, \avdist\
increases from 1 to 2$h$, or about 1.2 -- 2.5
kpc. These distances imply that the structure in \DMinf\ (and
\sigmaRM) is neither produced very close to the sun, nor outside the
Milky Way or its ionized halo.

For a line of sight at 20\degr\ from the Galactic plane, \avdist\ is
2$h$, or about 2.5 kpc. This is nearby enough that the truncation of
the Milky Way disk does not play a significant role in limiting the
calculated \DMinf. However, at smaller Galactic latitudes this can no
longer be ruled out, at 10\degr\ from the plane \avdist\ is already
4$h$ (5 kpc), and at 5\degr, \avdist\ = 8$h$ (10 kpc), so the model I
use to calculate \DMinf\ breaks down close to the Galactic plane. This
might be the reason why in Fig. \ref{sigma_rm.fig} \DMinf\ increases
much more rapidly at small latitudes than the measured \sigmaRM.




\section{The relative contributions of the Milky Way and extragalactic sources to \sigmaRM}\label{Sect.: ratio}\label{Sect.: The figure}
\begin{figure}
\resizebox{\hsize}{!}{\includegraphics{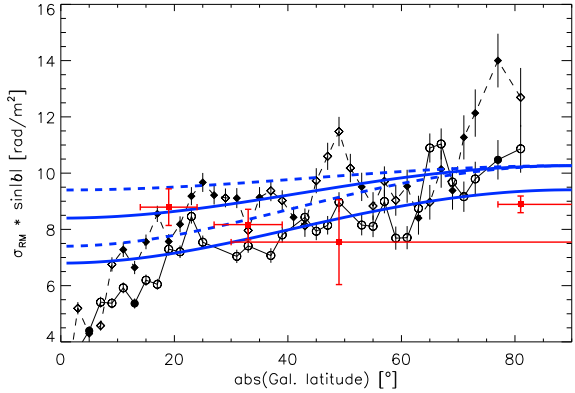}}

\caption{Distribution of \sigmaRM\ $\times$ sin$|b|$  as a function of Galactic latitude, for the data points shown in  Fig. \ref{sigma_rm.fig}. The data points at $\pm$ 1\degr\ and $\pm$ 3\degr\ are not shown; their \sigmaRM\ $\times$ sin$|b|$ are 2.0 and 4.5 resp. I fitted the model described in Sect. \ref{Sect.: ratio} to the data points at $|b|$ $>$ 20\degr\ for positive and negative latitudes separately, and show the best fits (solid  blue lines). To illustrate the sensitivity of the modelled \sigmaRM\ to $\sigma_{\mathrm{RM,MW}}$, I also draw 2 curves on either side of the best-fitting curve to the data at negative latitudes, that only differ in $\sigma_{\mathrm{RM,MW}}$ by 1 rad/m$^2$ from the best-fitting value. (dashed  blue lines) 
}
%
\label{sigma_avbparr.fig}
\end{figure}

Fig. \ref{sigma_avbparr.fig} shows \sigmaRM\ $\times$ sin$|b|$ as a function of Galactic latitude. 
The data points from Table \ref{sigmaRM.table} are nearly independent of latitude in this figure, which means that these \sigmaRM\ closely follow a 1/sin$|b|$ relation; the best-fitting DM$_\infty$ model from Sect. \ref{Sect.: ne_model} shows the same dependence on Galactic latitude. To explain why \sigmaRM\ and DM$_\infty$ might be correlated requires identifying which quantities determine how \sigmaRM\ varies with Galactic latitude.

The electron-density weighted line-of-sight component of the magnetic field, \avbparr, is defined as
\begin{eqnarray}
\langle B_{\|}\rangle \equiv\ \frac{\int\limits_{\mathrm{source}}^{\mathrm{observer}} n_{\mathrm{e}}\ B_\|\ \mbox{d}l}{\int\limits_{\mathrm{source}}^{\mathrm{observer}} n_{\mathrm{e}}\ \mbox{d}l}\ =\ \frac{\mathrm{RM}}{0.81\ \mbox{DM}_\infty} 
\nonumber
\end{eqnarray}
\noindent
 The total RM of the line of sight can then be
written in terms of its Galactic and extragalactic contributions as
\begin{eqnarray}
\mathrm{RM}\ =\ \mathrm{RM}_{\mathrm{MW}} + \mathrm{RM}_{\mathrm{EG}}\ =\ 0.81 \langle B_{\|}\rangle\ \mathrm{DM}_\infty + \mathrm{RM}_{\mathrm{EG}} 
\label{rm_los.eqn}
\end{eqnarray}
\noindent
with variance
\begin{eqnarray}
\sigma_{\mathrm{RM}}^2\ =\ 0.81^2\mathrm{DM}_\infty^2\sigma_{\langle B_\|\rangle}^2\ +\ 0.81^2\langle B_{\|}\rangle^2\sigma_{\mathrm{DM}_\infty}^2\ + \sigma_{\mathrm{RM,EG}}^2 
\label{sigmaRM.eqn}
\end{eqnarray}
\noindent
(assuming no correlation between the magnetic field and the free
electron density) In this Letter \sigmaRM\ is the (average) variance
of an ensemble of lines of sight within a single cell used to remove
the large-scale variation in $\langle\mathrm{RM}\rangle$ from the NVSS
data (Sect. \ref{Sect.: RM_lat}) or within the area covered by the
data sets from Table \ref{sigmaRM.table}.

A priori only $\sigma_{\mathrm{RM,EG}}$ is known to be independent of
Galactic latitude; the first two terms on the right-hand side of Eqn.
\ref{sigmaRM.eqn} can have both a $b$-dependent and a $b$-independent
part.  Although these different contributions to \sigmaRM\ could in
theory produce a complicated variation with latitude, the data show
that this behaviour can be modelled in a simple way, as \sigmaRM ($b$)
= $\sqrt{\left(
    \frac{\bar{\sigma}_{\mathrm{RM,MW}}}{\mathrm{sin}|b|}\right)^2 +
  \bar{\sigma}_{\mathrm{RM,EG}}^2}$. (I will use `$\bar{\sigma}$' to
distinguish the parameters from this model from the standard
deviations in Eqn. \ref{sigmaRM.eqn}) The model amplifies
$\bar{\sigma}_{\mathrm{RM,MW}}$, which itself does not depend on
Galactic latitude, at smaller Galactic latitudes as 1/sin$|b|$ to
mimic the increase in \DMinf\ at these latitudes, and then combines it
with $\bar{\sigma}_{\mathrm{RM,EG}}$ to form \sigmaRM. I show the best
fitting models in Fig. \ref{sigma_avbparr.fig} as blue lines, and
these fits reproduce the global behaviour of \sigmaRM\ with Galactic
latitude (for $|b|$ $\ge$ 20\degr). The parameters that produce these
fits at positive (/negative) latitudes are
$\bar{\sigma}_{\mathrm{RM,MW}}$ = 6.8 $\pm$ 0.1 (8.4 $\pm$ 0.1)
rad/m$^2$ and $\bar{\sigma}_{\mathrm{RM,EG}}$ = 6.5 $\pm$ 0.1 (5.9
$\pm$ 0.2) rad/m$^2$; the values for the reduced $\chi^2$ of these
fits are 4.4 and 3.9. I fitted the \sigmaRM\ at positive and negative
latitudes separately because they are off-set in
Fig. \ref{sigma_rm.fig}. To illustrate how sensitive the modelled
\sigmaRM\ are to different $\bar{\sigma}_{\mathrm{RM,MW}}$, I show for
negative latitudes also the best-fitting \sigmaRM\ models for
$\bar{\sigma}_{\mathrm{RM,MW}}$=9.4 rad/m$^2$ (top dashed curve) and
$\bar{\sigma}_{\mathrm{RM,MW}}$=7.4 rad/m$^2$ (bottom dashed
curve). Since the values of \sigmaRM\ $\times$ sin$|b|$ depend much
less on Galactic latitude for the data from table \ref{sigmaRM.table}
than for the NVSS RMs, their \sigmaRM\ can be modelled with a much
smaller $\bar{\sigma}_{\mathrm{RM,EG}}$ contribution.

Now the $\bar{\sigma}_{\mathrm{RM,MW}}$ and
$\bar{\sigma}_{\mathrm{RM,EG}}$ parameters from the model can be
identified with the terms in Eqn. \ref{sigmaRM.eqn}. The
$\mathrm{DM}_\infty\sigma_{\langle B_\|\rangle}$ and $\mathrm{\langle
  B_\|\rangle}\sigma_{\mathrm{DM}_\infty}$ terms can both have
$b$-independent parts, but since these terms add quadratically to
$\sigma_{\mathrm{RM,EG}}$, $\sigma_{\mathrm{RM,EG}}$ $\le$
$\bar{\sigma}_{\mathrm{RM,EG}}$. The $b$-dependent part of
Eqn. \ref{sigmaRM.eqn} is modelled as
$\bar{\sigma}_{\mathrm{RM,MW}}$/sin$|b|$. When all the structure of
the latitude-dependent part would be produced only by $\sigma_{\langle
  B_\|\rangle}$, which is equivalent to setting \sigmaDMinf=0, then
$\sigma_{\langle B_\|\rangle}$ =
$\bar{\sigma}_{\mathrm{RM,MW}}$/(0.81$\times$DM$_\infty$($b$=90\degr)).
\sigmaDMinf\ will however not be zero, since it contains a
contribution from the variation in modelled \DMinf\ over the region of
interest, and from local structure in the free electron density that
is not included in the smooth \DMinf\ model from Sect. \ref{Sect.:
  ne_model}. This structure in \DMinf\ can explain part of the
observed structure in \sigmaRM, and the calculated $\sigma_{\langle
  B_\|\rangle}$ is therefore an upper limit. (again since the $\sigma$
terms add quadratically) Using DM$_\infty$($b$=90\degr)=24.4
cm$^{-3}$pc, and an average value of $\bar{\sigma}_{\mathrm{RM,MW}}$ = 7.6
rad/m$^2$, then gives $\sigma_{\langle B_\|\rangle}$ $\le$ 0.4
$\mu$G. The \sigmaRM\ from Table \ref{sigmaRM.table} give the same
value when calculating \sigmaRM\ $\times$
sin$|b|$/(0.81$\times$24.4). (I ignored the $\sigma_{\mathrm{RM,EG}}$
term in this case -- please see the end of the previous paragraph)

The magnitude of the $\sigma_{\mathrm{RM,MW}}$ term compared to the
$\sigma_{\mathrm{RM,EG}}$ term implies that the Milky Way dominates
structure in RM on scales between about 1 -- 10 degrees, which can
affect studies of the RMs of extragalactic sources. (the lower limit
is set by the size scale probed by the NVSS RMs, and the upper limit
is set by the size of the bins used to subtract the variation in
$\langle$RM$\rangle$)

\section{Conclusions} In this Letter I show that the \sigmaRM\ of
  NVSS sources can be modelled as a dominant Galactic contribution
  ($\sigma_{\mathrm{RM,MW}}$ $\approx$ 8 rad/m$^2$) that is amplified
  at smaller Galactic latitudes as 1/sin$|b|$, similar to the increase
  in the Galactic free electron column density, \DMinf, and an
  extragalactic contribution that is independent of Galactic latitude
  ($\sigma_{\mathrm{RM,EG}}$ $\approx$ 6 rad/m$^2$). I corrected the
\sigmaRM\ of NVSS sources for the variation in $\langle$RM$\rangle$
along Galactic longitude, and for the broadening of the RM
distribution that is produced purely by the uncertainties in the NVSS
RMs.
The `excess' \sigmaRM\ correlates well with \DMinf. To calculate
\DMinf, I compared 4 models of the free electron density, and I
decided which one is the best based on how well it predicts the DM of
pulsars at known distances. This model builds up \DMinf\ on a
characteristic scale of a few kpc for latitudes $|b|$ $>$ 20\degr.  I
model the behaviour of \sigmaRM\ with Galactic latitude as an
extragalactic RM distribution, which does not depend on Galactic
latitude, and the Galactic RM contribution, which is amplified at
small Galactic latitudes to simulate the increase in \DMinf. The model
follows the observed global behaviour in \sigmaRM\ well, although
there are significant localized deviations between the model and the
observations. The best fitting values for the Galactic and
extragalactic contributions to RM for positive (/negative) Galactic
latitudes $\sigma_{\mathrm{RM,MW}}$ = 6.8 $\pm$ 0.1 (8.4 $\pm$ 0.1)
rad/m$^2$ and $\sigma_{\mathrm{RM,EG}}$ = 6.5 $\pm$ 0.1 (5.9 $\pm$
0.2) rad/m$^2$. By deriving which factors contribute to \sigmaRM\ I
show that this $\sigma_{\mathrm{RM,EG}}$ is an upper limit. The spread
in \avbparr\ that is required to produce a $\sigma_{\mathrm{RM,MW}}$ =
7.6 rad/m$^2$ is 0.4 $\mu$G.  However, fluctuations in the free
electron density will also contribute to $\sigma_{\mathrm{RM,MW}}$,
which means that this \sigmaAvbparr\ is an upper limit. I also
compiled \sigmaRM\ values from the literature, that have more accurate
RM determinations. These sources suggest that the
$\sigma_{\mathrm{RM,MW}}$ contribution is even larger, and that
$\sigma_{\mathrm{RM,EG}}$ is very small by comparison.

This result implies that structure in RM on angular scales between about
1\degr\ -- 10\degr is dominated by the Galactic foreground.

Future radio polarimetry surveys will provide accurate rotation
measures on a much finer grid on the sky. The analysis I present here
can be improved upon with such data sets; in particular they will
permit using smaller bins in Galactic longitude and latitude, which
can better correct for the variation in RM with Galactic longitude and
latitude.

\section*{Acknowledgements}
I would like thank Ettore Carretti, Bryan Gaensler, Jo-Anne Brown,
Neeraj Gupta, Ann Mao, Marijke Haverkorn, Peter Katgert, and Melanie
Johnston-Hollitt for their contributions at key points of this Letter,
Melanie Johnston-Hollitt and Naomi McClure-Griffiths for their careful
reading of the manuscript, and the anonymous referee for suggestions
that helped improve both the analysis and the manuscript.

\section*{Note added in press}
Since this paper was accepted it has become clear that the
$\sigma_{\mathrm{RM}}$ from Gaensler et al. and Johnston-Hollitt et
al. can contain significant contributions from their errors in RM,
which I should have taken into account. However, this does not affect
the conclusions from this Letter.

\end{document}